\documentclass[sigconf,review]{IEEEtran}

\usepackage{amssymb}

\usepackage{bm}
\usepackage{amsmath,amsfonts}
\usepackage{algorithm}
\usepackage{algorithmic}
\usepackage{graphicx}

\usepackage{subcaption}

\usepackage{textcomp}
\usepackage{xcolor}

\AtBeginDocument{%
  \providecommand\BibTeX{{%
    \normalfont B\kern-0.5em{\scshape i\kern-0.25em b}\kern-0.8em\TeX}}}

\begin{document}

\title{Approximate Wireless Communication for Federated Learning}

\author{\IEEEauthorblockN{Xiang Ma\IEEEauthorrefmark{1}, Haijian Sun\IEEEauthorrefmark{2}, Rose Qingyang Hu\IEEEauthorrefmark{1}, and Yi Qian\IEEEauthorrefmark{3}} \\
\IEEEauthorblockA{
\IEEEauthorrefmark{1}Department of Electrical and Computer Engineering, Utah State University, Logan, UT \\
\IEEEauthorrefmark{2}School of Electrical and Computer Engineering, University of Georgia, Athens, GA \\
\IEEEauthorrefmark{3}Department of Electrical and Computer Engineering, University of Nebraska-Lincoln, Lincoln, NE \\
Emails: \IEEEauthorrefmark{1}\{xiang.ma@ieee.org, rose.hu@usu.edu\}, \IEEEauthorrefmark{2}hun@uga,edu,  \IEEEauthorrefmark{3} yi.qian@unl.edu }
}
\maketitle

\begin{abstract}
This paper presents an approximate wireless communication scheme for federated learning (FL) model aggregation in the uplink transmission. We consider a realistic channel that reveals bit errors during FL model exchange in wireless networks. Our study demonstrates that random bit errors during model transmission can significantly affect FL performance. To overcome this challenge, we propose an approximate communication scheme based on the mathematical and statistical proof that machine learning (ML) model gradients are bounded under certain constraints. This bound enables us to introduce a novel encoding scheme for float-to-binary representation of gradient values and their QAM constellation mapping. Besides, since FL gradients are error-resilient, the proposed scheme simply delivers gradients with errors when the channel quality is satisfactory, eliminating extensive error-correcting codes and/or retransmission. The direct benefits include less overhead and lower latency. The proposed scheme is well-suited for resource-constrained devices in wireless networks. Through simulations, we show that the proposed scheme is effective in reducing the impact of bit errors on FL performance and saves at least half the time than transmission with error correction and retransmission to achieve the same learning performance. In addition, we investigated the effectiveness of bit protection mechanisms in high-order modulation when gray coding is employed and found that this approach considerably enhances learning performance.
\end{abstract}


\maketitle

\section{Introduction}
Federated learning (FL) \cite{konevcny2016federated} enables local devices to perform machine learning (ML) tasks while still benefiting from the learning generalization ability provided by model parameter sharing. It does not require sharing locally collected data among devices and the server. Instead, only model parameters are shared, thereby effectively protecting data privacy. The FL system is composed of a central parameter server (PS) and a large number of smart local clients (LCs). LCs gather data from onboard sensors and execute a predefined ML task based on the global model broadcast from the PS. After computation, each LC sends local models to the PS for aggregation, and then the updated global model is redistributed back to LCs. This process is repeated until the global model converges. 

For edge devices such as UAVs serving as LCs, wireless networks are usually employed to connect them to the PS. However, the nature of wireless channels often results in erroneous information transmission. To address this issue, modern wireless communications utilize forward error correction (FEC) methods, such as convolutional code and low-density parity check code (LDPC), to detect and correct received bit errors. The basic principle of FEC is to encode the message with redundant information in the form of an error correction code (ECC). The receiver can correct the error bits without knowing the actual bits sent by the transmitter. Packet retransmission can be employed when the number of errors exceeds the correction capability of ECC. Although FEC and packet retransmission are powerful, they increase computation and communication overhead, leading to extra power consumption and transmission delays during FL model aggregation. In \cite{shirvanimoghaddam2022federated}, the authors focused on transmission bit errors in FL but only in a packet erasure channel.

Stochastic gradient descent (SGD) is a widely used optimization method in distributed ML. In FL, each client performs SGD on ML tasks, then a single-step gradient is calculated and sent to the central PS in every communication round. This method, called FedSGD \cite{mcmahan2017communication}, serves as a baseline algorithm for FL. However, for large-scale distributed ML models with millions of parameters, transmitting gradients can cause high delay. Advanced transmission schemes such as non-orthogonal multiple access (NOMA) \cite{9207963} are good options, but they need to equip with complex decoding methods. Gradient compression is a promising approach to addressing this challenge, where extensive research has shown the effectiveness of gradient sparsification and quantization with little performance loss. For instance, 1-bit SGD was applied in \cite{seide20141} to reduce gradient transmission size, and in \cite{aji2017sparse}, it was shown that 99\% of gradients could be dropped. Therefore, we are motivated to apply \emph{approximate wireless communication} to transmit those gradients, i.e., allowing lossy transmission (with errors) in exchange for low latency, low overhead, and less FEC computation in this paper. We would like to emphasize that gradient compression is different from and runs parallel to our proposed approximate wireless communication method, even though both can result in information errors. The tolerance of gradient quantization errors is based on the assumption that the gradient magnitude is small enough. Empirical studies in \cite{wen2017terngrad, m2021efficient, guo2021partition} have shown that the gradients are close to Gaussian distribution, and most gradient values fall in the range of $(-1, 1)$ or even $(-0.01, 0.01)$. Approximate wireless communication for media data transmission has been proposed in \cite{ransford2015sap, sen2010design}, and similar ideas can be applied to FL gradient transmission.

In this study, we present a theoretical analysis of bounded ML gradients in commonly used ML settings. Specifically, we prove that gradients in fully connected neural network models and convolutional neural network models are bounded under commonly used conditions. Based on this analysis, we set a limit on the erroneous gradient, together with the approximate transmission in practical wireless networks. Simulation results demonstrate that our proposed method is effective in reducing the impact of bit errors on FL performance, and saves half the time than transmission with Error Correction and ReTransmission (ECRT). The rest of this paper is organized as follows. Section 2 introduces the FL model in wireless networks, Section 3 presents the theoretical analysis of bounded gradients, Section 4 describes the proposed method, Section 5 presents simulation results, and Section 6 concludes the paper.

\section{System Model}
The FL model is considered as follows. The FL system consists of $M$ LCs, which are connected to the PS through wireless channels. The overall data amount $D$ is distributed among $M$ devices, with each device $m$ containing $D_m$ data.  

\subsection{FL System Model}
FL is an iterative ML algorithm that performs local computation and global aggregation in each round. Local computation is followed by the LC model uploading, and after global aggregation, the global model is downloaded to each LC. This process is repeated until the model converges. The objective function of FL can be defined as: 
\begin{equation}
    \min_{\bm{w} \in R^d} f(\bm{w}) \quad  \text{where} \quad f(\bm{w}) \stackrel{\text{def}}{=}\frac{1}{|D|}\sum_{i=1}^{|D|} f_i(\bm{w}), \label{eq:objective}
\end{equation}
where $|D|$ is the size of dataset $D$. $f_i(\bm{w}) = C(\bm{x_i}, \bm{y_i};\bm{w})$ is the cost or loss function used to measure the inference error between the data sample $(\bm{x_i}, \bm{y_i})$ and the inference made by model parameters $\bm{w}$. For classification problems in ML, the cross-entropy function is commonly used as the loss function $C$, particularly in neural network models. In multiclass classification, the label $\bm{y_i}$ is typically one-hot encoded to ensure each label carries equal weight.

As the data is distributed among $M$ LCs and not on the same device, the objective function (\ref{eq:objective}) needs to be rewritten as follows:
\begin{equation}
    f(\bm{w}) = \sum_{m=1}^M \frac{|D_m|}{|D|} F_m(\bm{w}), 
\end{equation}
where $F_m(\bm{w}) = \frac{1}{|D_m|}\sum_{i\in D_m} f_i(\bm{w})$. By distributing the data and computation across multiple devices, a conventional centralized ML problem can be transformed into a distributed FL problem.

Since the cost function for neural networks is typically non-convex, it is challenging to solve directly and find the global minimum. Therefore, the gradient descent method is an iterative optimization algorithm commonly used in ML to find a local minimum point. Stochastic gradient descent (SGD) is a variant of the gradient descent method that can be helpful in escaping local minimums by selecting data samples randomly. As a result, gradients play a central role in the learning process. The gradient is defined as:
\begin{equation}
    g = \nabla_{\bm{w}} C(\bm{x_i}, \bm{y_i};\bm{w}).
\end{equation}
The local gradient at each LC in each round can be written as 
\begin{equation}
    g_t^m = \nabla F_m(w_t).
\end{equation}
And the global gradient after aggregation is
\begin{equation}
    g_t = \sum_{m=1}^M \frac{|D_m|}{|D|} g_t^m.
\end{equation}
The PS stores the model weights from the last round $w_{t}$, and then updates the global model as follows
\begin{equation} \label{eq:sgd_update}
w_{t+1} = w_{t}-\eta g_t.    
\end{equation}
Here, $\eta$ is the learning rate, which typically falls within the range of $(0,1)$. 

\subsection{Wireless Channel Model}
Federated Learning is an upper-layer algorithm that does not have knowledge of the lower-layer gradient transmission details. Typically, transmission takes place over wireless channels when LCs are smart sensors or UAVs. For the uplink channel from LCs to PS, we consider the fading channel, which can lead to random bit errors. For the downlink channel, we assume the PS can deliver global gradients to LCs with negligible errors, this can be justified by higher PS transmit power (hence higher SNR)~\cite{shirvanimoghaddam2022federated}.

In the uplink, a time division scheme can be used where each user is assigned to a specific time slot while sharing the same channel. The received signal at the PS  can be expressed as follows
\begin{equation}
    r_t^m = \sqrt{p_t^m (d^m)^{-\alpha} } h_t^m g_t^m + n_t^m,
\end{equation}
where $r_t^m$ represents the signal received at the PS from client $m$. The transmission power is denoted as $p_t^m$, and the small scale fading is denoted as $h_t^m$, which is assumed to be complex normal Gaussian distributed, i.e., $h_t^m \sim \mathcal{CN} (0, 1)$. The distance between the PS and client $m$ is represented by $d^m$, and the path-loss exponent is denoted as $\alpha$. The additive noise is given as $n^t \sim \mathcal{CN} (0, \sigma^2)$. PS has the knowledge of the channel gain, i.e., $c_t^m=\sqrt{p_t^m (d^m)^{-\alpha} } h_t^m$, and only the noise serves as an error source.

The entire transmission process can be described as follows. First, the gradients are converted from decimal format to binary format. The bits are then mapped to symbols using a QAM modulation scheme. The symbols are then transmitted through the wireless fading channel. At the receiver end, the signal is decoded with maximum likelihood estimation and then demodulated to the closest point in the constellation, 
\begin{equation}
    \hat{g}_t^m = \arg_{\Bar{g}_t^m \in \mathcal{G}} \min || r_t^m - \sqrt{p_t^m (d^m)^{-\alpha} } h_t^m \Bar{g}_t^m||^2,
\end{equation}
where $\mathcal{G}$ is the symbol points set of the constellation diagram.

\section{Bounded Gradients under Constraints - A Sketch of Proof }


\subsection{Gradient Backpropagation}
In machine learning, particularly in deep neural networks, backpropagation is widely used for calculating gradients in each layer. In a fully connected neural network, the feed-forward equation at each neuron can be expressed as
\begin{equation}
\begin{aligned}
& z_j^l = b_j^l + \sum_k w_{jk}^l a_{k}^{l-1}, \\
& a_j^l  = \sigma(z_j^l).
\end{aligned}    
\end{equation}
Here, $b$ is the bias, $w$ is the weights, $z$ is the intermediate output, and $a$ is the final output after the activation function $\sigma(\cdot)$. $l$ represents $l$-th layer and $j, k$ are indices. The corresponding four fundamental equations in back-propagation for a fully connected network is 
\begin{subequations}
\begin{alignat}{2}
& \delta_j^L = \frac{\partial C}{\partial z_j^L} = \frac{\partial C}{\partial a_j^L} \frac{\partial a_j^L}{\partial z_j^L} = \frac{\partial C}{\partial a_j^L}  \sigma'(z_j^L) \label{eq:bp1},\\
& \delta_j^l = \frac{\partial C}{\partial z_j^l} = \sum_k \frac{\partial C}{\partial z_k^{l+1}} \frac{\partial z_k^{l+1}} {\partial a_j^l} \frac{\partial a_j^l}{\partial z_j^l} = \sum_k \delta_k^{l+1} w_{kj}^{l+1}\sigma'(z_j^l) \label{eq:bp2},\\
& \frac{\partial C}{\partial b_j^l} = \frac{\partial C}{\partial z_j^l} \frac{\partial z_j^l}{\partial b_j^l} = \delta_j^l \label{eq:bp3},\\
& \frac{\partial C}{\partial w_{jk}^l}  = \frac{\partial C}{\partial z_j^l} \frac{\partial z_j^l}{\partial w_{jk}^l} = \delta_j^l a_k^{l-1} \label{eq:bp4}.
\end{alignat}
\end{subequations}
Here, $L$ is the final layer index in the neural network, and $\delta_j^l$ is the defined ``error'' in $l$-th layer at node $j$. 

To ensure that the gradient $\nabla C = \frac{\partial C}{\partial w_{jk}^l}$ is bounded, it is necessary to limit $\delta_j^l$ and $a_k^{l-1}$ based on equation (\ref{eq:bp4}). These two terms are discussed separately. $a_k^{l-1}$ is the activation function output of neuron $k$ at the $(l-1)$-th layer. It depends on the activation function being used. For example, the Sigmoid function ensures that $a_k^{l-1}$ is in the range $(0, 1)$ regardless of the input $z_k^{l-1}$, while the ReLU activation function requires the input to be bounded. Further discussion and mathematical expressions on activation functions can be found in \cite{nwankpa2018activation}.

The calculation of $\delta_j^l$ is described in equation (\ref{eq:bp2}), which involves a summation of products of next layer errors $\delta_k^{l+1}$, weights from node $j$ in the $l$-th layer to next layer $w_{kj}^{l+1}$, and the derivative of activation function $\sigma'(z_j^l)$. There are three terms in this equation, and the summation requires the number of neurons in each layer to be finite. The derivative of the activation function $\sigma'(z_j^l)$ also depends on the activation function used, with the derivative being in the range $(0, 0.25)$ for the Sigmoid function and $\{0,1\}$ for ReLU. The weight $w_{kj}^{l+1}$ depends on model initialization, learning rate $\eta$, and last round gradient based on equation (\ref{eq:sgd_update}). Weight initialization methods typically generate random weight values in the range $(-1, 1)$ or even smaller, and there are newer initialization methods such as \cite{glorot2010understanding} and \cite{he2015delving}. Without loss of generality, we assume that the weight value $w_{kj}^{l+1}$ is bounded. The error $\delta_k^{l+1}$ can be written in the same way as in equation (\ref{eq:bp2}) with elements in the $(l+2)$-th layer, and this process continues all the way back to the final layer. In classification problems, the softmax function is commonly used as the activation function in the final layer to normalize the output class probabilities. When the cross-entropy loss function is used, it can be combined with the softmax function. For cross-entropy loss function, 
\begin{equation}
    C = -\sum_i  y_i log(p_i),
\end{equation}
where $y_i$ is the input truth label, $p_i$ is the softmax probability for the $i$-th class
\begin{equation}
    p_i = \sigma(z_i) = \frac{e^{z_i}}{\sum_k e^{z_k}},
\end{equation}
and the derivative is 
\begin{equation}
\frac{\partial p_i}{\partial z_j} = \left \{
  \begin{aligned}
    & p_i (1-p_j), && \text{if}\ i = j; \\
    & -p_j \cdot p_i, && \text{if}\ i \neq j. 
  \end{aligned} \right.
\end{equation}
Equation (\ref{eq:bp1}) can be written as 
\begin{equation}
\begin{aligned}
\delta_j^L = \frac{\partial C}{\partial p_i^L} \frac{\partial p_i^L}{\partial z_j^L} & = -\sum_i y_i \frac{\partial log(p_i)}{\partial p_i} \frac{\partial p_i}{\partial z_j}, \\
        & = -\sum_i y_i \frac{1}{p_i} \frac{\partial p_i}{\partial z_j}, \\
        & = -y_j(1-p_j) - \sum_{i \ne j} y_i \frac{1}{p_i}(-p_j \cdot p_i), \\ 
        & = p_j \cdot \sum_i y_i - y_j.
\end{aligned}
\end{equation}

Since $y$ is a one-hot encoded label vector, so $\sum_i y_i=1$, that is
\begin{equation}
    \delta_j^L = p_j - y_j.
\end{equation}
As $p_j$ takes values between 0 and 1 and $y_j$ is either 0 or 1, $\delta_j^L$ lies in the interval $(-1,1)$. 

To summarize, in a fully connected neural network with cross-entropy as the cost function and softmax function as the activation function in the final layer, the final layer error $\delta_j^L$ is in the range $(-1,1)$. In addition, if the weights are assumed in the range $(-1,1)$ and Sigmoid functions are used as activation functions in other layers, the gradient $\frac{\partial C}{\partial w_{jk}^l}$ is bounded by the sum of the number of neurons after $l$-th layer, denoted as $B^l$.

\subsection{Gradient in Convolutional Neural Network}
Modern image recognition tasks often use convolutional neural networks (CNNs) as an advanced technique. CNNs are a special variant of feedforward networks that consist of three types of layers: convolutional layers, pooling layers, and fully connected layers. The feedforward process of a CNN can be written as:
\begin{subequations}
\begin{alignat}{2}
& z_{j,k}^1  = b_{j,k}^1 + \sum_p \sum_q w_{p,q}^1 x_{j+p, k+q}^0 \label{eq:cnn_conv1}, \\
& a_{j,k}^1  = \sigma(z_{j,k}^1) \label{eq:cnn_conv2}, \\
& a_{j,k}^2  = \max(a_{2j,2k}^1, a_{2j+1,2k}^1, a_{2j,2k+1}^1, a_{2j+1,2k+1}^1) \label{eq:cnn_pool},\\
& z_i^3  = b_i^3 + \sum_{j,k} w_{i;j,k}^3 a_{j,k}^2 \label{eq:cnn_fc1},\\
& a_i^3  = \sigma(z_i^3) \label{eq:cnn_fc2}.
\end{alignat}
\end{subequations}
For the sake of simplicity, we assume that this CNN network comprises only three layers. Equation (\ref{eq:cnn_conv1}) and (\ref{eq:cnn_conv2}) represent the convolutional layer, equation (\ref{eq:cnn_pool}) represents the max pooling layer with a $2\times 2$ kernel, and equations (\ref{eq:cnn_fc1}) and (\ref{eq:cnn_fc2}) represent the fully connected layer. Here, $p$ and $q$ denote the indices of convolutional kernels.

Now the backpropagation for the CNN network becomes
\begin{subequations}
\begin{alignat}{2}
 \delta_i^3 &= \frac{\partial C}{\partial z_i^3} = \frac{\partial C}{\partial a_i^3}\frac{\partial a_i^3}{\partial z_i^3} = \frac{\partial C}{\partial a_i^3 }\sigma'(z_i^3) \label{eq:cnn_bp1},  \\
 \delta_{j,k}^1 &= \frac{\partial C}{\partial z_{j,k}^1} = \sum_i \frac{\partial C}{\partial z_{i}^3} \frac{\partial z_{i}^3}{\partial a_{s,t}^2} \frac{\partial a_{s,t}^2}{\partial z_{j,k}^1} \label{eq:cnn_bp2}, \\ &= \sum_i \delta_i^3 w_{i;s,t}^3 \frac{\partial a_{s,t}^2}{\partial a_{j,k}^1} \frac{\partial a_{j,k}^1}{\partial z_{j,k}^1},\nonumber \\
& = \sum_i \delta_i^3 w_{i;s,t}^3 \frac{\partial a_{s,t}^2}{\partial a_{j,k}^1} \sigma'(z_{j,k}^1),  \nonumber \\
& = \begin{cases}
 \sum_i \delta_i^3 w_{i;s,t}^3 \sigma'(z_{j,k}^1), &\text{if case 1;} \\
0, &\text{otherwise;} 
\end{cases} \nonumber \\
 \frac{\partial C}{\partial w_{i;j,k}^3} &= \frac{\partial C}{\partial z_{i}^3} \frac{\partial z_{i}^3}{\partial w_{i;j,k}^3} = \delta_i^3 a_{j,k}^2 \label{eq:cnn_bp3}, \\
 \frac{\partial C}{\partial w_{p,q}^1} &= \frac{\partial C}{\partial z_{j,k}^1} \frac{\partial z_{j,k}^1}{\partial w_{p,q}^1} = \delta_{j,k}^1 x_{j+p, \ k+q}^0.\label{eq:cnn_bp4}
\end{alignat}
\end{subequations}
Here, case 1 is $a_{j,k}^2 = \text{max}(a_{2s,2t}^1, \ a_{2s+1,2t}^1, \ a_{2s,2t+1}^1,\ a_{2s+1,2t+1}^1)$ in equation (\ref{eq:cnn_bp2}), and $s=\frac{j}{2}, n = \frac{k}{2}$.

Similarly, when the cross entropy serves as the loss and the activation function in the last layer is the softmax function, $\delta_i^3$ lies in the range $(-1,1)$. In order to bound the gradient $\frac{\partial C}{\partial w_{i;j,k}^3}$ in the fully connected layer, it is necessary to ensure that $a_{j,k}^2$ or $\max(a_{2j,2k}^1, a_{2j+1,2k}^1, a_{2j,2k+1}^1, a_{2j+1,2k+1}^1)$ is also bounded. If the Sigmoid function is used as the activation function in the first layer, then $a_{j,k}^2$ is bounded within the range of $(0,1)$, which results in $\frac{\partial C}{\partial w_{i;j,k}^3}$ being bounded within the range of $(-1,1)$. The gradient $\frac{\partial C}{\partial w_{p,q}^1}$ in equation (\ref{eq:cnn_bp4}) can be bounded by considering the two terms involved. Firstly, $x_{j+p, k+q}^0$ is the input and is bounded. Secondly, $\delta_{j,k}^1$ can be expressed as either 0 or $\sum_i \delta_i^3 w_{i;s,t}^3 \sigma'(z_{j,k}^1)$. We know that $\delta_i^3$ is in the range $(-1,1)$ and $\sigma'(z_{j,k}^1)$ is in the range $(0,0.25)$ if the activation function is Sigmoid. For $w_{i;s,t}^3$, its value depends on the model initialization, learning rate $\eta$, and the last round gradient based on equation (\ref{eq:sgd_update}), as we discussed above. If we assume $w_{i;s,t}^3$ is bounded in $(-1,1)$, then $\frac{\partial C}{\partial w_{p,q}^1}$ can be bounded by the number of neurons in the last layer, also denoted as $B^l$.

\section{Proposed Method}
In section 3, we have presented mathematical proofs that under certain conditions, the gradients are bounded by $B^l$. Empirically, it has been shown in \cite{wen2017terngrad, m2021efficient, guo2021partition} that the gradients are not only bounded but also bounded within the range of $(-1, 1)$ or even a smaller range. This allows us to have an expected value (statistically) for the received gradient at the PS. Correspondingly, we first design a QAM encoding scheme for the bounded gradients, 


\subsection{QAM Encoding}
In ML, gradients are commonly expressed using 32-bit floating-point numbers. These numbers follow the format defined by the IEEE-754 standard, which assigns the first bit to the sign and the next 8 bits to the exponent, leaving the final 23 bits for the fraction. Bits in different locations have varying importance. The sign bit controls the sign of the gradient value, while the exponent part defines the integer and decimal values. The fraction part only controls the decimal value, and thus, the exponent bits are more important than the fraction bits. Furthermore, the bits located on the left side of the exponent are more important than those on the right.

\begin{figure}[ht]
	\includegraphics[width=3.0in]{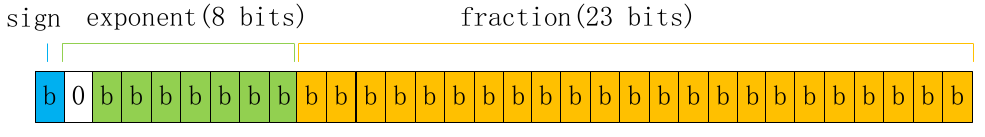}
	\centering
	\caption{Received Gradient Bit Representation}
	\label{received_bit}
	\centering
\end{figure}

During transmission, each bit is susceptible to noise, which can cause corruption. To avoid block corruption, we employ interleaving at the transmitter and de-interleaving at the receiver, reducing the likelihood of multiple error bits taking place together. In the bit representation, when the second bit in the 32-bit representation, i.e., the first bit in the exponent part, is 1 and all other 31 bits are 0s, the decimal value is 2. Conversely, when the second bit in the 32-bit representation is 0, and all other 31 bits are 1s, the magnitude is less than 2. When assuming a magnitude threshold of 1 for the gradient value, the first bit in the exponent part is always 0. This motivated us, on the receiver side, regardless of the value decoded in the second-bit location of the gradient, it will be set to 0, as shown in Figure \ref{received_bit}.

Moreover, we have also observed that different modulation schemes have varying effects on bits located at different positions \cite{sen2010design}. This is important not only in media message transmission but also in ML model parameter transmission. In wireless transmission, the transmission system is not aware of the relative importance of data bits and treats all the bits equally. When using QPSK as the modulation scheme, each symbol consists of 2 bits, with the bit combinations being ${00, 01, 11, 10}$. The error probability for the first and second bits in QPSK is the same. In contrast, 16-QAM has 4 bits per symbol, and the constellation map with gray code is shown below.
\begin{figure}[ht]
	\includegraphics[width=1.5in]{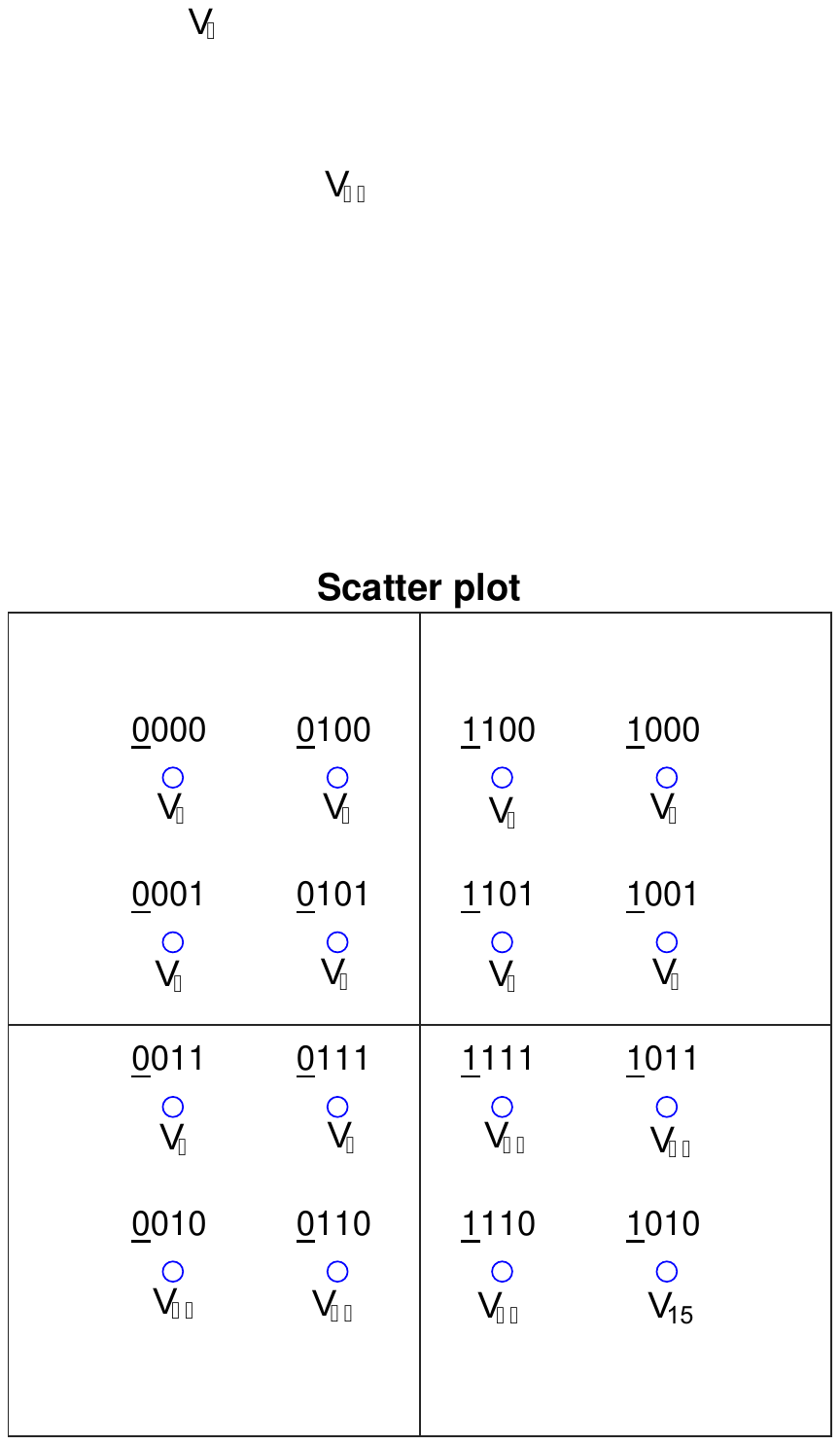}
	\centering
	\caption{16-QAM with Gray Coding Constellation Map}
	\label{qam16_map}
	\centering
\end{figure}

The bits with underlines in Figure \ref{qam16_map} correspond to the first bit in each symbol, which are the most significant bits (MSB), while the fourth or last bits are the least significant bits (LSB). When the transmission probability for each symbol is the same, the error probability for the MSB is higher than for the LSB. For example, if the symbol $s_0$ is decoded with an error, it is most likely to be decoded as $s_1$, $s_4$, or $s_5$. The MSB bit remains the same, while the LSB changes twice. This is summarized in Table \ref{Tab:error_count}. The symbols in the other quadrants are symmetric to the first quadrant, so the results are identical. High-order modulation schemes with gray coding provide built-in protection for the MSB bits of gradient values in bit representation.
\begin{table}[h]
	\newcommand{\tabincell}[2]{\begin{tabular}{@{}#1@{}}#2\end{tabular}}
	\centering
	\caption{16-QAM MSB/LSB Error Count}
	\begin{tabular}{c|p{32mm}|p{15mm}|p{14mm}}
		\hline
		\textbf{Symbol} & \textbf{Potential Error Symbol} & \textbf{MSB Error Count} & \textbf{LSB Error Count} \\
		\hline
		$s_0$ & \tabincell{l}{$s_1, s_4, s_5$} & 0 & 2 \\
		\hline
		$s_1$ & \tabincell{l}{$s_0, s_2, s_4, s_5, s_6$} & 2 & 3\\
		\hline
		$s_4$ & \tabincell{l}{$s_0, s_1, s_5, s_8, s_9$} & 0 & 2\\
		\hline
            $s_5$ & \tabincell{l}{$s_0, s_1, s_2, s_4, s_6, s_8, s_9, s_{10}$} & 3 & 3\\
		\hline
	\end{tabular}
	\label{Tab:error_count}
\end{table}

\subsection{Approximate Wireless Transmission}
To further improve gradient exchange efficiency over wireless networks, we propose an approximate wireless transmission scheme. Essentially, since the gradient is resilient to errors, as witnessed in the existing gradient compression methods, the delivered messages (gradient) do not have to be accurate. Hence we eliminate FEC and re-transmission when channel SNR is satisfactory.  While the exact SNR value is to be determined, our empirical results have shown that at around 10-20 dB, the BER is acceptable for FL. Notice that our approach is different from user datagram protocol (UDP), where retransmission is not required either. The difference is that UDP works at a higher level, and the CRC is used only to check the UDP payload. When the error happens at the physical or MAC layer, retransmission is still issued.  Our approach eliminates both FEC and re-transmission at lower layers, including physical and MAC layers.  The benefit is three-fold: 1) it reduces the communication overhead, so more data bits can be transmitted; 2) it reduces the computation overhead for FEC, and this is very appealing for edge devices; 3) it improves latency performance since no re-transmission is required.

\section{Simulation Results}
In this section, we first present the parameter settings for the simulation. We then provide the empirical probability of bit error versus signal-to-noise ratio (SNR) under the wireless channel mentioned earlier. Among the modulation schemes tested, QPSK achieves a better bit error rate (BER) than 16-QAM and 256-QAM at the same SNR level. Next, we compare the FL performance under three scenarios: ECRT transmission with error correction and retransmission, naive erroneous transmission, and erroneous transmission with our proposed scheme. The naive error transmission is the transmission in wireless networks with errors without extra operation. Compared to naive erroneous transmission, our proposed scheme achieves a high testing accuracy. Furthermore, compared to ECRT transmission, our proposed scheme for erroneous transmission saves much more time. Finally, we discuss different modulation schemes with gray coding to show the built-in bit protection for MSB in the bit representation.

We consider a typical FL setting in our simulation, where $M=100$ LCs are connected to the PS, and all LCs participate in the learning process in each communication round. The LCs perform image classification tasks using the MNIST dataset, which consists of handwritten digits 0-9. The training set contains $60,000$ images, and the test set contains $10,000$ images, with each digit having approximately $6,000$ images in the training set and $1,000$ images in the test set. To simulate a realistic scenario where data is collected from the environment, we distribute the data in a non-iid way, with each LC having $2$ digits and each digit having around $300$ images for training. We use a convolutional neural network (CNN) as the ML model, with 2 convolutional layers, each having a kernel size of 5, 2 max-pooling layers with size 2, and 2 fully connected layers. ReLU is used as the activation function in all layers except the last one, which uses the log softmax function. The learning rate is set to $\eta=0.01$.

We set the path loss exponent for the wireless channel as $\alpha=3$, and consider a distance of 10m between the PS and LCs. The transmission power at the LCs is normalized to 1. We use QPSK as the modulation scheme, and the receiver SNR is set at $\gamma=10$ dB unless otherwise specified.

Under the specific fading channel, QPSK achieves a lower BER compared to 16-QAM and 256-QAM at the same SNR level. For QPSK, at SNR=10 dB, the BER is approximately $4 \times 10^{-2}$ while the BER is $5 \times 10^{-3}$ when SNR is 20 dB.

\begin{figure}[ht]
	\includegraphics[width=2.8in]{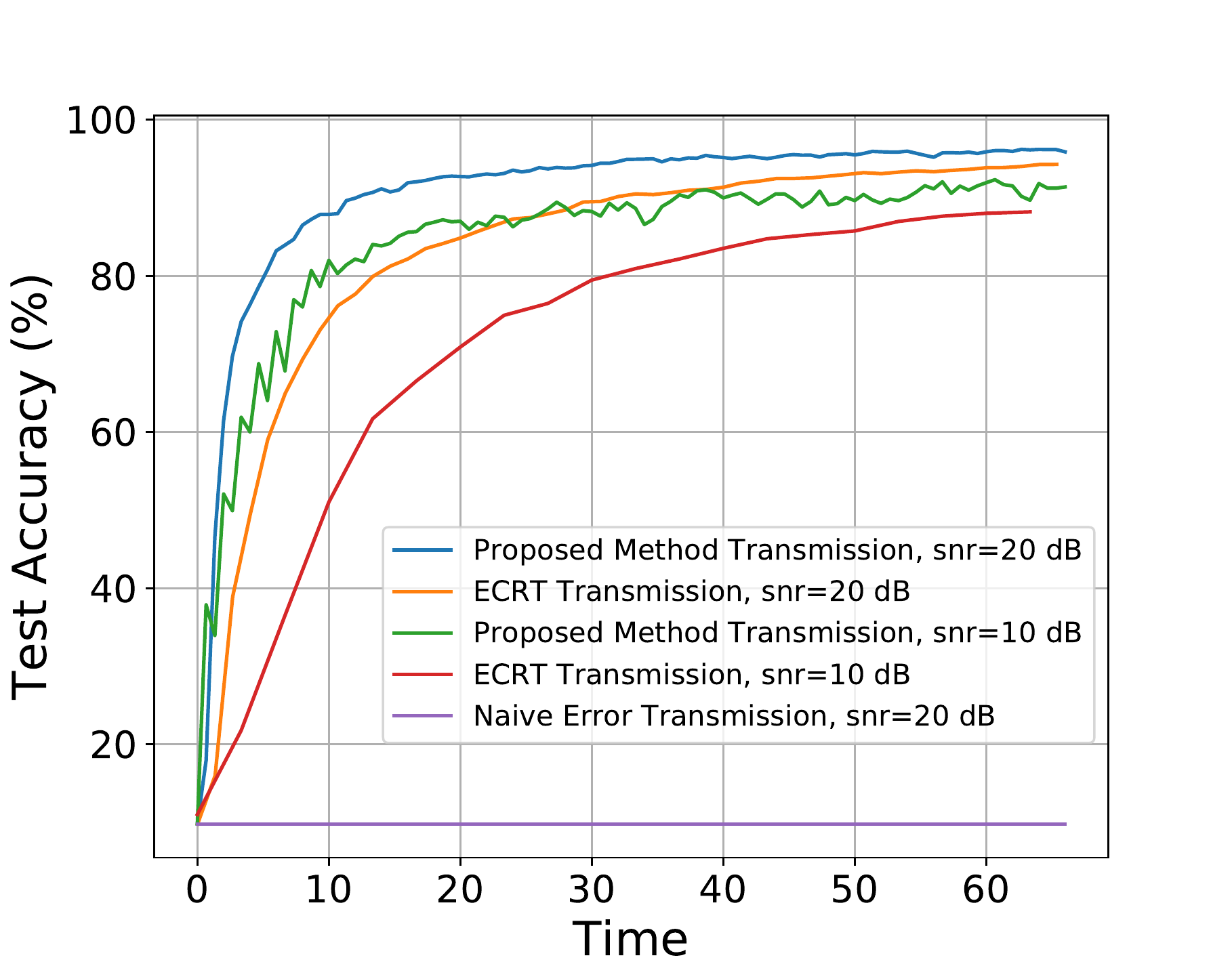}
	\centering
	\caption{Test Accuracy v.s. Communication Time}
	\label{mnist_niid_time}
	\centering
\end{figure}

In the ECRT scheme with error correction and retransmission, all the bits are received correctly by the PS, which incurs a cost for forward error correction (FEC) and possible retransmission when the error exceeds the FEC capability. In contrast, the naive error transmission scheme involves transmitting bits with errors without prior knowledge of the gradients, where the test accuracy remains flat at around 10\%, similar to random guessing as shown in Figure \ref{mnist_niid_time}. This occurs because the model cannot learn anything due to transmission errors. Our proposed method, however, takes into account prior knowledge of the gradient values, which are expected to be in the range of (-1, 1). This makes the proposed scheme achieves much better results than naive error transmission.  

To quantify the transmission time saved by our proposed method compared to ECRT transmission, we employ a practical IEEE 802.11 protocol with LDPC error correction coding. LDPC is a promising ECC that can approach the Shannon limit. For different coding rates, there exists a trade-off between error correction capability and transmission overhead. Lower coding rate results in high transmission overhead but comes with high error correction capability. Here, we use a coding rate of $1/2$ to enhance error correction. According to \cite{butler2016minimum}, the minimum Hamming distance is 15 for a code rate of $1/2$ when the code length is 648, and we search using the parity check matrix. This results in an error correction capability of 7 bits. In Figure \ref{mnist_niid_time}, the transmission with LDPC coding with retransmission takes $2\times$ time than the proposed scheme to achieve 80\% accuracy at SNR=20 dB while it takes more than $3\times$ for SNR=10 dB for the LDPC coding with retransmission scheme to achieve that performance.

\begin{figure}[ht]
  \centering
  \subfloat[Test Accuracy with the Same SNR=10 dB]{\includegraphics[width=0.23\textwidth]{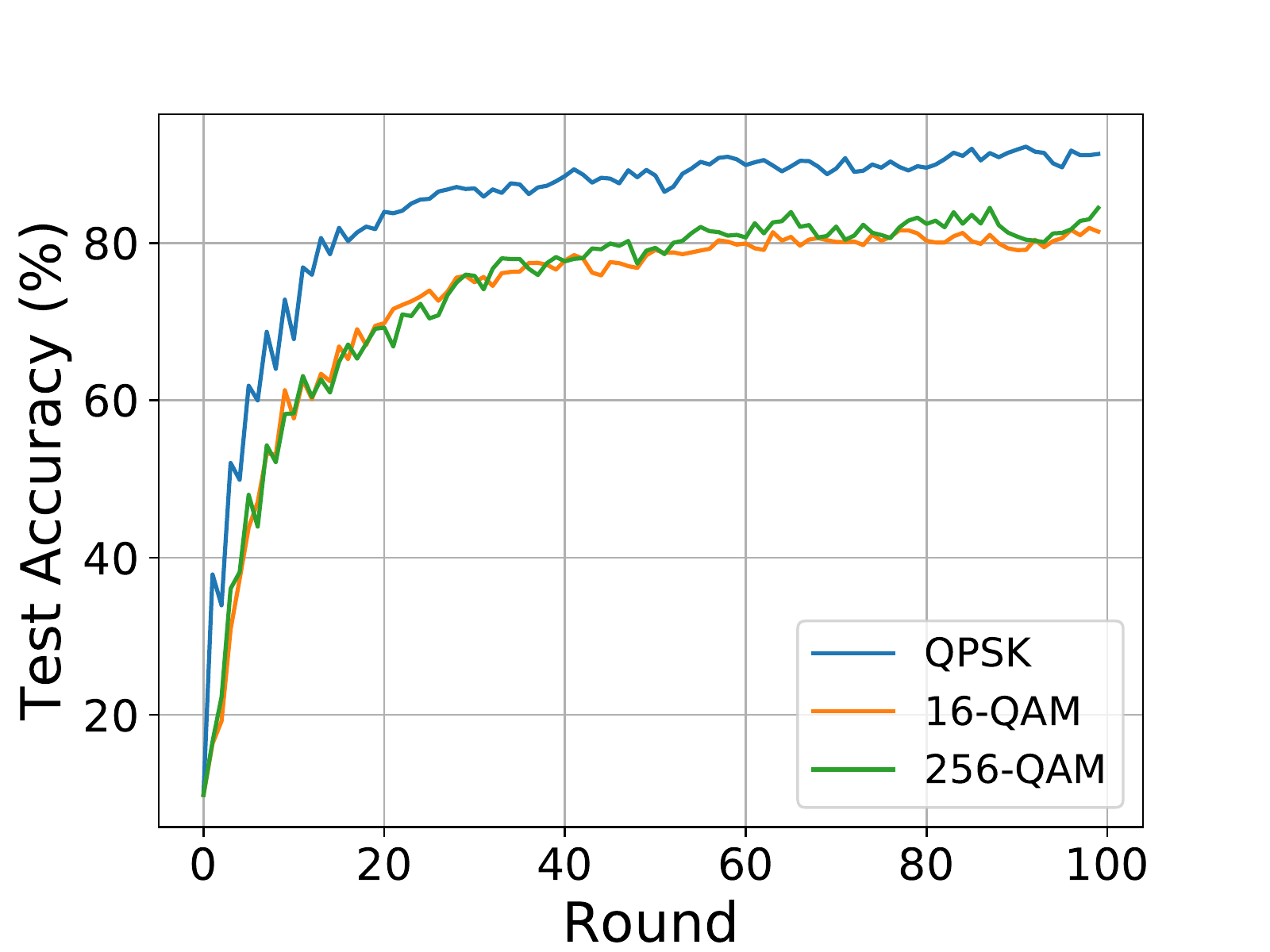}\label{fig:f1}}
  \hfill
  \subfloat[Test Accuracy with the Same BER $\approx 4 \times 10^{-2}$]{\includegraphics[width=0.23\textwidth]{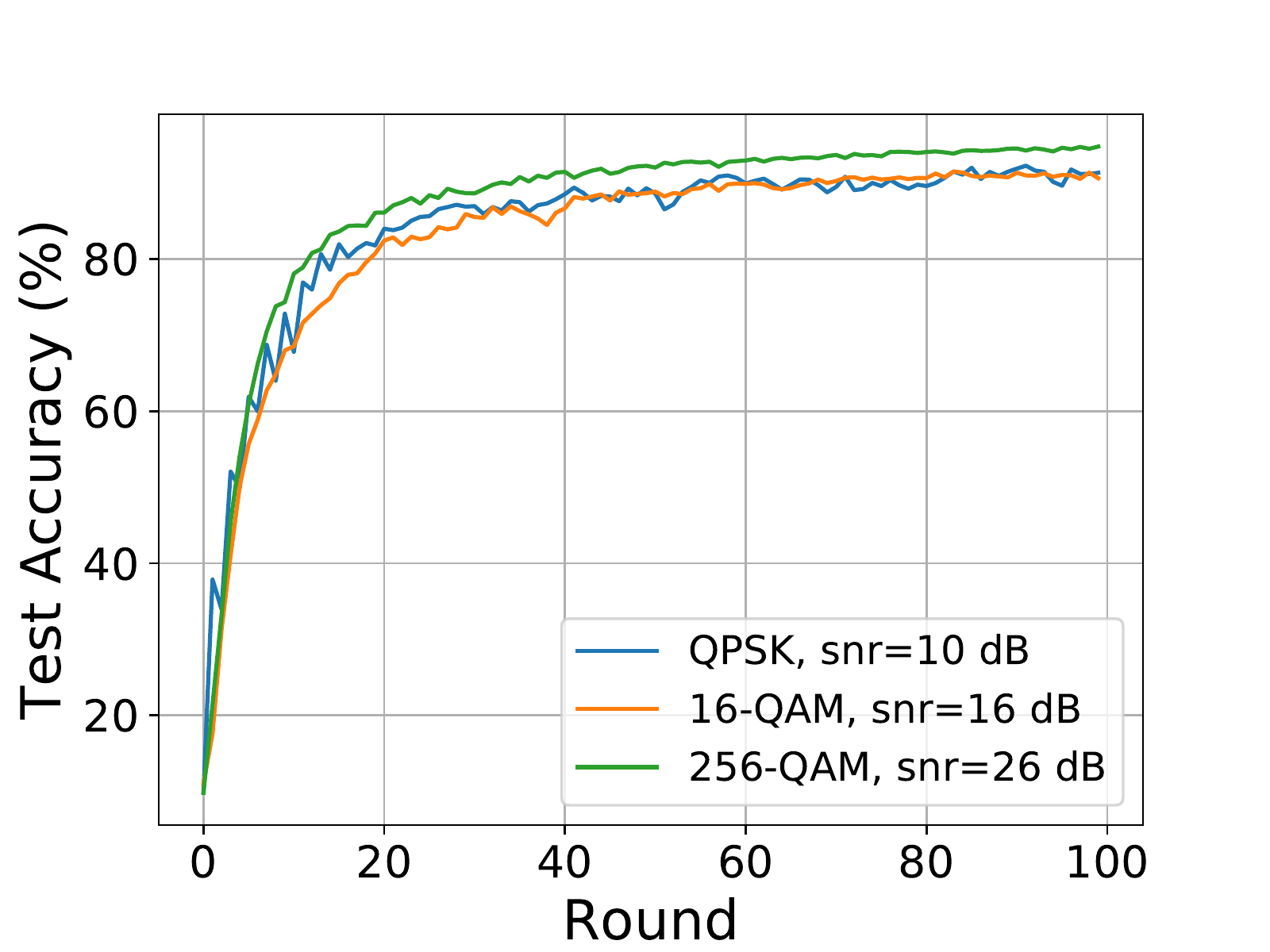}\label{fig:f2}}
  \caption{Test Accuracy with the Same SNR/BER}
  \label{same_ber_snr}
\end{figure}

To demonstrate the effectiveness of built-in MSB bit protection of high-order modulation with gray coding, we begin by presenting the test accuracy of different modulations at the same SNR in Figure \ref{same_ber_snr}(\subref{fig:f1}). At an SNR of 10 dB, the BER for QPSK, 16-QAM, and 256-QAM is roughly $4 \times 10^{-2}$, $10^{-1}$, and $3 \times 10^{-1}$, respectively. Because QPSK results in fewer errors, its learning performance is better than in 16-QAM and 256-QAM.


In Figure \ref{same_ber_snr}(\subref{fig:f2}), we present a scenario where the BER is made the same for different modulations. To accomplish this, we increase the SNR for 16-QAM to 16 dB and the SNR for 256-QAM to 26 dB. Consequently, the BER for all three modulation schemes is $4 \times 10^{-2}$. In this scenario, 256-QAM achieves significantly better learning performance than QPSK, with smaller transmission errors in 256-QAM than QPSK.

\section{Conclusions}
In this paper, we proposed a federated learning parameter transmission scheme in wireless networks. Unlike existing transmission methods that rely on forward error correction and retransmission, we proposed gradient transmission with errors based on prior knowledge of gradient values. The gradient value is mathematically proven to be within a small range under certain constraints, so the received gradient value is expected to be within that range. This approach achieves learning performance with errors much better than naive error transmission and saves at least half time to achieve the same learning performance as ECRT transmission. Additionally, we explored high-order modulation and demonstrated improved learning results. In the future, our plan is to quantify the impact of communication errors on FL performance and examine how the number of clients influences it.


\end{document}